 \documentclass[12pt, preprint]{aastex}  
 \shorttitle{} \shortauthors{Domingue et al.}  
\begin{document}   
 \title{2MASS/SDSS Close Major-Merger Galaxy Pairs: Luminosity Functions and Merger Mass Dependence}  
 \author{Donovan L. Domingue\altaffilmark{1}}
 \affil{Department of Chemistry and Physics, Georgia College and State University, CBX 082, Milledgeville, GA 31061} 
\email{donovan.domingue@gcsu.edu}  
\and  
\author{C. K. Xu, T. H. Jarrett and Y. Cheng\altaffilmark{1}}
 \affil{Infrared Processing and Analysis Center, California Institute of Technology, MS 100-22, Pasadena, CA 91125 } 
\email{cxu@ipac.caltech.edu, jarrett@ipac.caltech.edu,even@ipac.caltech.edu}    
\altaffiltext{1}{Visiting Astronomer , Kitt Peak National Observatory, National Optical Astronomy Observatory, which is operated by the Association of Universities for Research in Astronomy, Inc. (AURA) under cooperative agreement with the National Science Foundation. }  
 \begin{abstract} 

We select a close ``major-merger candidate" galaxy pair sample in order to calculate the $K_{s}$ luminosity function (LF) and pair fraction representative of the merger/interaction component of galaxy evolution in the local universe. The pair sample (projected separation 5 h$^{-1}$ kpc $\leq$ r $\leq$ 20 h$^{-1}$ kpc, $K_{s}$-band magnitude difference $\Delta$$K_{s}$ $\leq$ 1 mag) is selected by combining the Two Micron All Sky Survey (2MASS) with the Sloan Digital Sky Survey (SDSS) Data Release 5 (DR5). The resulting data set contains 
340 galaxies covering 5800 sq. degrees. A stellar mass function is also translated from the LF. A differential pair fraction displays nearly constant fraction of galaxy pairs as a function of galaxy mass from 10$^{9}$ to 10$^{11.5}$ M$_{\sun}$ . The differential pair fraction is less subject to absolute magnitude bias due to survey limitations than the standard total pair fraction.  These results suggest that major-merger candidate pairs in the 0$<$z$<$0.1 universe are developed from $\sim$1.6$\%$ of the galaxy population without dependance on galaxy mass for pair components below 10$^{11}$ M$_{\sun}$. The derived LF combined with merger model time scales give local merger rates per unit volume which decrease with masses greater than 10$^{11}$ M$_{\sun}$ .
\end{abstract}  

 \section{Introduction}
 Galaxy-galaxy interactions/mergers play a central role in the important processes in galaxy evolution, including mass assembly, star formation, morphological transformation, and AGN activity. Merging of galaxies can lead to the formation of larger galaxies and galaxy structures (Kauffmann, White, \& Guiderdoni 1993; Cole et al. 2001). Theoretical simulations (Barnes 1990) and observations (Schweizer 1982; Kormendy \& Sanders 1992) show that gas-rich late-type galaxies transform to gas-poor early-type E/S0 galaxies through galaxy mergers. The central black holes in active galactic nuclei (AGNs) are likely to be built up mostly in galaxy mergers, given the tight correlation between the black hole mass and the bulge mass (Franceschini et al. 1999). Tidal torques produced during the interaction may send a large amount of gas into the galactic nuclear region and feed a preexisting black hole, leading to enhanced AGN activity (Domingue, Sulentic, \& Durbala 2005) and enhanced SFR as a function of galaxy separation (Geller et al. 2006), in particular for close major mergers (Xu \& Sulentic 1991).
 
 There is strong evidence that galaxy-galaxy interactions/mergers can significantly enhance the star formation rate (SFR) in galaxies involved (see Kennicutt 1998 for a review). Galaxies with similar mass, i.e. ``major interactions", are more likely to develop enhanced star formation than ``minor interactions" (Daysra et al. 2006; Woods, Geller, \& Barton 2006). Evolution in the cosmic SFR is likely due to a population of peculiar/interacting starburst galaxies that are closely related to galaxy mergers (Brinchmann et al. 1998) although alternative scenarios have been presented (Bell et al. 2005; Melbourne et al. 2005; Faber et al. 2005). Pairs may even include 
 $\sim$ 50\% of luminous galaxies at z=1 to 3 (Kartaltepe et al. 2007). Major mergers may also have contributed to the current state of $\sim$50\% of present-day massive galaxies (M$^{*}$$>$5$\times$10$^{10}$ M$_{\sun}$; Bell et al. 2006).

     Identification of galaxy mergers/interactions can be done by two methods including the selection of binary galaxies and identifying galaxies with peculiar morphology. Studies based on the latter method find strong evolution in the fraction of major-mergers (galaxy pairs with mass ratio $<$3), particularly for massive galaxies (Brinchmann et al. 1998; Le F\'{e}vre et al. 2000; Conselice et al. 2003). However, these results have significant uncertainties because it is difficult to quantify the morphological peculiarity. In contrast, it is easy to define binary galaxies quantitatively and objectively. This makes an objectively defined comparison between local merger events and high-z merger events possible. However, earlier studies of pair fraction and its cosmic evolution have suffered seriously from the contamination of unphysical pairs because of the lack of redshifts or highly incomplete redshift data (Zepf \& Koo 1989; Burkey et al. 1994; Carlberg, Pritchet, \& Infante 1994; Yee \& Ellington 1995; Woods, Fahlman, \& Richer 1995; Patton et al. 1997; Wu \& Keel 1998). In recent studies using samples of galaxies with measured redshifts, Le F\'{e}vre et al. (2000) and Patton et al. (2002) found m = 2.7 $\pm$ 0.6 and 2.3 $\pm$ 0.7, respectively, where m is the evolution index in the power law fit ($\sim$ (1+z)$^{m}$) of the merger rate as a function of redshift, z. Photometric redshifts were also used to identify pairs (Kartaltepe et al. 2007) in the COSMOS field and these authors find evidence for stronger evolution with m = 3.1 $\pm$ 0.1. In a series of papers, Patton et al. (1997, 2000, 2002) pointed out that in studies of merger rate and evolution, it is very important to control various systematic biases, otherwise results from comparisons between mismatched samples of low-z and high-z galaxies are not very meaningful.

     The merger rate and its cosmic evolution can be constrained with comparison of differential pair fraction functions at different redshifts. A differential pair fraction function (DPFF) is defined by ratios between the number of paired galaxies and that of all galaxies in luminosity bins (Xu, Sun, \& He 2004; XSH). Such functions are not sensitive to sample selection (flux limited or volume limited) and therefore can be compared without bias between different studies. DPFF can be determined by comparing the luminosity (mass) function of paired galaxies with that of total galaxies. We estimate the local Ks (2.16 $\mu$m) band luminosity function (LF) of close major-merger pairs and derive from it the DPFF in the 0$<$z$<$0.1 universe. The close relation between the Ks-band luminosity and the stellar mass allows us to generate the mass function of the paired galaxies and the mass dependence of the merger rate. Since this can be compared directly to the predictions of the hierarchical galaxy formation simulations (e.g., Benson et al. 2002), it will provide an important test for these simulations. A small pair sample (19 pairs) from XSH and the implications contained therein that the DPFF is mass dependent motivates a new determination of DPFF from the larger redshift data set in the Sloan Digital Sky Survey (SDSS) Data Release 5 (DR5; Adelman-McCarthy et al. 2007).
 We adopt the $\Lambda$-cosmology with $\Omega$$_{m}$=0.3, $\Omega$$_{\Lambda}$=0.7, and h=H$_{0}$/(100 km s$^{-1}$ Mpc$^{-1}$).
 
 \section{Sample Selection} 
Pair candidates are selected from a parent sample which was created by matching the Sloan Digital Sky Survey (SDSS) Data Release 5 (DR5; Adelman-McCarthy et al. 2007) spectroscopic galaxy catalog with the Two Micron All Sky Survey (2MASS) Extended Source Catalog (XSC; Jarrett et al. 2000). The selected DR5 galaxy spectroscopic sample represents 561,530 quality redshift measurements which matched a sample of 70,126 unique galaxies in the XSC. In order to determine the redshift coverage of our parent sample we make the matched galaxies a subset of the XSC with a limiting magnitude of $K_{s}$ $<$ 13.5, the completeness limit of the XSC (Jarrett et al. 2000), and use this catalog as our basis for the parent sample. The default value of $K_{20}$ is used for the $K_{s}$-band magnitude (Jarrett et al. 2000).
From this catalog, the parent sample is restricted to galaxies which have a redshift completeness index c$_{z}$$>$0.5, where c$_{z}$ is the ratio of the number of galaxies with measured redshifts within 1\degr $ $ radius from the center of the galaxy in question and the number of all galaxies within the same radius. The parent sample meeting this criteria consists of 77,451 galaxies, of which 66,478 have measured redshifts (86\% redshift completeness). 

All galaxies with redshift in the parent sample are candidate primary galaxies around which we search for neighbors by modifications of the methods of XSH. Neighbors are not required to have a measured redshift. When neighbors are found, they must meet the following criteria to be included as pair members: (1) the $K_{s}$ magnitude of the primary is not fainter than 12.5. (2) At least one component must have a measured redshift. (3) If both components have measured redshift, the velocity difference is not larger than 1000 km s$^{-1}$. (4) The projected separation is in the range of 5 h$^{-1}$ kpc $\leq$ r $\leq$ 20 h$^{-1}$ kpc. When only one component has a measured redshift, the separation is calculated according to that redshift and the angular separation of the components. (5) The $K_{s}$ difference between the two galaxies is not larger than 1 mag. Criteria (1) and (5) require all selected galaxies to be brighter than $K_{s}$ = 13.5, ensuring the completeness of the pair sample. Criteria (3) has a velocity separation which is larger than that imposed in XSH and Patton et al. (2000). The added population of pairs with a wider velocity separation range of 500 km s$^{-1}$ $\leq$$\Delta$v$\leq$ 1000 km s$^{-1}$ comprise $\leq$9\% of the population with 2 measured redshifts (Fig. 1). We justify the inclusion of this population in order to retain physical pairs in environments of higher density and velocity dispersion. 
Contamination by unphysical pairs in this range should not be significant (Patton et al. 2000). Criteria (4) and (5) imply that the selected pairs will be ``close major-merger pairs". These criteria reduce the contamination of unphysical pairs among those with one redshift measurement. ``Major-merger" is defined here as a pair with a mass ratio no greater than 2.5, which is selected by criteria (5).  Pair selection is not restricted on the basis of local galaxy density. Analysis of pair populations based on local environment is presented in section 4.

Photometry derived from standard 2MASS pipeline can be uncertain for pairs closer than 30$\arcsec$.
 One of us, THJ, has deblended the K-band magnitudes for close pairs by means of galaxy profile fitting and subtraction. These newly derived magnitudes are used throughout this work. Geller et al. (2006) find 12\% of their CfA2 pairs to be unresolved in 2MASS. For these pairs with a separation of near 10$\arcsec$ or smaller, the 2MASS XSC automatic photometry will detect the brighter galaxy and include close companions as part of the photometric magnitude with no dimmer counterpart in the XSC. This introduces a population of missed pairs in a 2MASS selected sample for which it is necessary to visually inspect and perform more sophisticated photometry routines to de-blend the two close companions. Since the pair criteria restrict our sample to pairs with a projected separation r $\geq$ 5 h$^{-1}$ kpc, the closest galaxy pairs can only become closer than 10$\arcsec$ beyond z=0.034. The primary galaxy candidate list contains 8837 galaxies at z$\geq$0.034. After inspection of the candidates, 126 potential pairs were selected for further photometric analysis for the magnitude criteria (1) and (5) required for inclusion in our sample.
Of these 126 pairs, 51 are determined to be additional ``close major-merger pairs"  and added to our
automated selection data set.

A search through published redshifts from NASA Extragalactic Database (NED) for the selected pairs adds 58 redshift measurements to the sample. Of these new measurements, 13 pairs are excluded by selection criteria (3) (22\% unphysical). 
In order to decrease the contribution of unphysical pairs to the portion of our sample with single redshifts, three nights of spectroscopy at the KPNO 2.1m were completed. Redshifts for 22 pair members were acquired (Table 1) and 4 pairs were shown to be unphysical (18\%). These new redshifts results combined with the literature search suggest the the contribution of unphysical pairs among the single redshift sample is indeed $\sim$20\%.
Among the selected sample, 7 have a z$\geq$0.1 mostly contributed (6 of 7) by the added 2MASS unresolved pairs. To avoid contamination by evolution with redshift and in order to have a well defined redshift range, we remove these 6 pairs from our final analysis samples. The final sample contains 170 pairs (340 galaxies). Among these pairs, 122 have both components with measured redshifts, and 48 are single redshift pairs. 

\section{Observations and Data Reduction}

We obtained three nights at the KPNO 2.1m to perform spectroscopic observations of pairs with single known redshifts. The time alloted allowed us to obtain redshifts for 22 galaxies. The observations were carried out with the 2.1m GoldCam spectrograph with grating 32 for a dispersion of 2.47 \AA / pixel.
The spectra were reduced with the standard NOAO IRAF packages for bias removal, flat fielding and sky subtraction and wavelength calibration. The IRAF package FXCOR was used to obtain redshifts utilizing the absorption lines between 4000-7000 \AA. The newly acquired galaxies with redshifts are listed in Table 1 with an indication of inclusion or rejection as a pair member after obtaining this new data. 

\section{Pair Sample Characteristics}

Redshift and $K_{s}$ magnitude distribution of the parent and pair sample are given in Figure 2.
The pair sample is complete to $K_{s}$ = 12.5 because of selection criteria (1) while the parent sample exhibits the same completeness level as XSC (complete to $K_{s}$ = 13.5). This completeness difference also affects the distribution of redshift between the parent and pair samples (Fig.2). 
Examination of the portion of the two samples with $K_{s}$ $<$12.5 reveals that the pair sample has retained a fairly representative magnitude and redshift distribution as compared to its parent sample, with a range 0.001 $<$ z $<$ 0.10. The median redshift of the pair sample is z=0.04.
The $K_{s}$-band selection criteria employed here are known to bias toward red sequence galaxies with larger mass (Obri\'{c} et al. 2006) near M$^{*}$. The SDSS redshift survey is limited to r $<$ 17.6. So galaxies with $K_{s}-r$$>$ 4.1 may be missed in our $K_{s}$-selected sample, introducing a color-related incompleteness. This effect should be insignificant because in the local universe these galaxies are very rare.
Peculiar velocity corrections (Mould et al. 2000) are applied to the entire sample which corrects for the effect of the Local Group, Virgo, Virgo Great Attractor, and Shapley infall velocities, and therefore luminosity uncertainties are minimized. The conversion to absolute magnitudes are determined with, 

\begin{equation}
M_{K}= K_{20}-25-5log(D_{L})-k(z), 
\label{eq1}
	\eqnum{1}
\end{equation}
where D$_{L}$ (Mpc) is the luminosity distance from our adopted $\Lambda$-cosmology, and k(z) is the k-correction; k(z)=-6.0log(1+z). This k-correction utilized by Kochanek et al.(2001) is valid for z$<$0.25 and is independent of galaxy type.

Galaxy and pair morphology can indicate the history of pair formation as compared to galaxies in the field. The color $u-r$=2.22 separates late-type spirals (Sb, Sc, Irr) from early-type (E,S0, Sa) galaxies (Strateva et al. 2001). The color comparison samples for both the pairs and a selected control sample  are dominated by early-types. As an isolated galaxy control sample, we select galaxies from the 2MASS/SDSS parent sample of redshift completeness index c$_{z}$$>$0.5,  with all neighbors of magnitude difference $\Delta$$K_{s}$ $\leq$ 2 mag, having a distance from the galaxy,r $\geq$ 500 h$^{-1}$ kpc. The restriction on neighbor magnitude difference limits the completeness to galaxies with $K_{s}$ $\leq$ 11.5. Only the galaxies with $K_{s}$ $\leq$ 11.5 in both the pairs and control galaxy samples are compared for color differences. This sample comparison contains 217 control galaxies and 59 paired galaxies. The control galaxy sample ($K_{s}$ $\leq$ 11.5) is composed of 95.4\% red sequence and the comparison paired galaxy sample ($K_{s}$ $\leq$ 11.5) is 88.1\% red sequence. 
Obri\'{c} et al. (2006) used a 2MASS/SDSS matched catalog from SDSS DR1 to find that their combined catalog has 80\% red galaxies vs. 66\% for all SDSS galaxies. Using this $u-r$ criteria our final paired galaxy sample consists of 85\% red seqeunce (E,S0, Sa) which is comparable to the value of the matched DR1 catalog (Obri\'{c} et al. 2006). An alternate set of galaxy types assigned based on the median type determined visually by two of the authors (DD \& KX) along with an automated classification with E/S0 having an inverse concentration index c $<$ 0.35 (Shimasaku et al. 2001, Strateva et al. 2001) and SDSS color u-r $>$ 2.2, reduces the fraction of E/S0/Sa galaxies in our sample to near 50\%. This alternate typing scheme reveals that many of the spirals in both comparison and our pair sample have red colors. The red population of Sloan survey galaxies indeed has been shown to contain a significant fraction of spirals (Lintott et al. 2008). In lieu of acquiring a set of visual types from the large 2MASS/SDSS parent sample, we retain the $u-r$ criteria to investigate pair morphology with the caveats that types listed here are more an indication of color morphology than physical type.

Pair morphology can be assigned to three types as (1) those consisting of two blue sequence galaxies (S+S), (2) pairs with two red sequence galaxies (E+E), and (3) so called mixed morphology pairs consisting of one of each type (E+S). The pair formation scenario should influence the fraction of each pair type.  Random combinations of these field galaxies without environmental variation in their distribution would generate 32\% E+S when the field is $\sim$80\% E/S0 as found in Obri\'{c} et al. (2006) . Also, an alternative  scenario with local environment as the only factor to determine pair formation would dictate very low numbers of mixed E+S pairs as the Holmberg (1958) effect would be expected to create galaxies of similar type within that environment. The addition of secular evolution based on galaxy-galaxy interactions and mergers (``nurture") can change galaxy types and increase the numbers of E+S in the local environment (``nature'') scenario (Domingue et al., 2003 ; Junqueira, de Mello, \& Infante 1998). 
The observed numbers of E+S pairs are near $\sim$19\% in our total, isolated, and grouped sample. 
This is below the number of purely random capture scenario, and results in the E+E fraction $\sim$0.75 which is above the 0.64 predicted by random combinations of this field. The Holmberg effect here is seen in red sequence pairs while S+S pairs are represented in the fraction $\sim$0.04-0.05 which is closer to expected from random combinations. This is an indication of the strong role of environment and possibly mergers in the history of red sequence galaxies.
  
Isolation of the pairs and the environmental effects on pair LF are investigated by dividing the pair sample into an isolated and ``grouped" pair sample. The isolated pairs are designated as such when all neighbors with a magnitude difference $\Delta$$K_{s}$ $\leq$ 2 mag, have a distance from the pair center,r $\geq$ 100 h$^{-1}$ kpc, or have a velocity difference $\Delta$$v$$>$10$^{3}$ km s$^{-1}$.
All other pairs are considered to be members of triplets, groups or clusters and are assigned as ``grouped" pairs. Assigning pair environment in this manner yields 103 isolated pairs and 67 ``grouped" pairs for our sub-samples.

 \section{Pair Fraction and $K_{s}$-Band Luminosity Function}
Biases present in the pair sample include both missed and unphysical pairs. The minimum fiber separation of the SDSS spectroscopic observations ($\sim$55\arcsec ; Blanton et al. 2003a) is a systematic source for missing pairs.
Our requirement that only one galaxy of the pair must have a redshift reduces this source of incompleteness. In a purely 2D search for pair candidates in the parent sample, 2152 galaxies have neighbors closer than 55\arcsec.
Of these, only 244 (11\%) are in 'pairs' having no redshift measurement for either component.
We use this estimate to statistically determine the effect of missed pairs on our pair fraction and luminosity functions.

The second bias of contamination by unphysical pairs, mostly from the single redshift pairs can be estimated by Monte Carlo simulations of XSH(2004). However, because it is assumed in the simulations that galaxies are uniformly distributed (XSH),  they may have underestimated  the number of unphysical pairs by neglecting the clustering effects. We know from a literature search with NED that 10 of 48 ``literature double redshift" pairs are unphysical. We therefore adopt a more conservative estimate of 20\% $\pm$ 20\% (10 $\pm$ 10) as the likely contribution of unphysical pairs to the single redshift sample.

The likelihood of being a false pair is proportional to the searching area, which is inversely proportional to $z^{2}$, multiplied by $n$ where $z$ and $n$ are respectively, the redshift and local density (r$\leq$ 10\arcmin) of  neighboring galaxies of $\vert K_{s}-K_{s}^{'}\vert$ $\leq$ 1. This is in the following ``false factor":

\begin{equation}
  Q_{false,i}=\left\{
	\begin{array}{ll}
		0 \pm 0 
		\mbox{   (2 redshifts pairs),}  \\
		(10 \pm 10)(n_{i}/z_{i}^{2})/\sum_{j}(n_{j}/z_{j}^{2}) &\\
		 \mbox{(1 redshift pairs),}	
	\end{array}\right.
	 \label{eq2}
		 \eqnum{2}
\end{equation}

where the summation is over the 48 single redshift pairs.

The pair fraction can be estimated as:
\begin{equation}
	f_{p}=\frac{A}{N_{g}}\sum_{i}^{N_{pg}}(1-Q_{false,i}),
	\label{eq3}
	\eqnum{3}
\end{equation}

where $N_{g}$= 17,793 is the total number of galaxies in the parent sample brighter than $K_{s}$ = 12.5 with 0$<$z$<$0.1, $A$= 1/(1-0.11) is the correction factor to compensate for missing pairs, and
 $N_{pg}$ = 265
 is the total number of galaxies in the pair sample brighter than $K_{s}$ = 12.5. The error of $f_{p}$, err = $(A/N_{g})$$\{$$\sum_{i}^{N{pg}}$$[(1-Q_{false,i})^{2}+e_{Q,i}^{2}]\}^{1/2}$, where $e_{Q}$ is the error in equation (2).
Using these formula, we find a pair fraction in equation (3), $f_{p}= 1.6\% \pm 0.1\%$.

\subsection{1/V$_{max}$ Luminosity Function}
The first method  we use to find the $K_{s}$-band LF of paired galaxies is the 1/V$_{max}$ method (Schmidt 1968).
The effective sky coverage is determined by comparing the 2MASS number counts of our parent sample to those of Kochanek et al. (2001). We estimate this sky coverage as 5800 deg$^{2}$ with an error of $\sim$3\%. Given our selection criteria,
 both pair components have the same V$_{max}$ determined by the redshift of the pair, the $K_{s}$ magnitude of the primary, and the limiting magnitude; $K_{lim}$=12.5. The LF and its error are calculated as in XSH(2004) by the following formulae:

\begin{equation}
\phi(M_{K.i})=\frac{A}{\delta(m)}\sum_{j}^{N_{i}}\frac{1-Q_{false,j}}{V_{max,j}},
\eqnum{4}
\end{equation}
\begin{equation}
e_{\phi}(M_{K,i})=\frac{A}{\delta(m)}\sqrt{\sum_{j}^{N_{i}}\frac{(1-Q_{false,j})^{2}+e_{Q,j}^{2}}{V_{max,j}^{2}}},
\eqnum{5}
\end{equation}
where $\phi(M_{K,i})$ is
the LF in the $i$th bin of the $K_{s}$-band absolute magnitude, $N_{i}$ is the number of galaxies in that bin, $\delta(m)=0.5$ is the bin width, and V$_{max,j}$ is the maximum finding volume of the $j$th galaxy in the bin. Other symbols are defined by equation (3). The results for the LF of the entire pair sample, the isolated pair sample and the pairs in groups/clusters are listed in Tables 2-4 and plotted in Figure 3-4. The parameters for the best fitting Schechter functions (Schechter 1976) are given in Table 5.

Stellar masses, corresponding to the absolute magnitude bins, are also listed in Table 2-4. The isophotal $K_{s}$ magnitude is translated to the ``total" $K_{s}$ magnitude ($\Delta K_{s}$ =0.2 mag).
Following XSH (2004), Kochanek et al. (2001), and Cole et al. (2001), we assume a conversion factor of $M_{star}/L_{K}$ = 1.32 $M_{\sun}/L_{\sun}$ which is from a Salpeter initial mass function (Cole et al. 2001). The differential pair fractions (Tables 2-4, Figures 3-4) are calculated using the Schechter functions of the paired galaxy samples and of 2MASS galaxies (Kochanek et al. 2001). The error is estimated from the quadratic sum of the error of the LF of paired galaxies and its deviation from the Schechter function. Bins with 1 galaxy or less are not included in the Schechter fit or the differential pair fraction plot due to their large uncertainty.

\subsection{Maximum Likelihood Luminosity Function}
It is well known that inhomegeneous distribution of galaxies such as clustering can affect the LF derived from the  1/V$_{max}$-method therefore, a step-wise maximum likelihood method (SWML) developed by Efstathiou, Ellis, \& Peterson (1988; EEP) for determining the LF is also applied to the pair data. This method has many advantages including (1) it is insensitive to possible inhomogeneous distributions of galaxies in the sample and (2) it is independent on the true and assumed functional form being the same (i.e. the Schechter form).  
The SWML method is applied here as in EEP with the modification that each galaxy is weighted according to its probability of being a physical pair, (1- $Q_{false,i}$), where $Q_{false,i}$ is given in equation (2) and we have selected the calculation to include the same number of bins as 1/V$_{max}$ calculations. The SWML implementation was done with the public software package KCORRECT (v4.1.4)(Blanton et al. 2003b). The calculated SWML points are included in Fig. 3-4 as a comparison to the 1/V$_{max}$. The normalization is chosen to match the 1/V$_{max}$ LF in the most populated bin which includes M$^{*}$. With this alternate LF we verify the shape of the 1/V$_{max}$ LF. The Schechter fit and the SWML LF differ the most at $M_{K}$$\sim$ -24 suggesting that the true pair LF at the bright end may not be of the exact Schechter form.

\subsection{Sample Selection Biases and Environmental Effects}
By selecting randomly placed galaxies within the parent sample and pairing them into 200 synthesized pairs with the same magnitude and redshift requirements as our pair sample, we test for any LF bias introduced by our pair selection criteria and choice of parent sample. To avoid synthetic pair samples with an overabundance of single redshift pairs due to the lack of position constraints, galaxies without measured redshift are excluded. The galaxies are grouped into 200 random pairs for 100 simulations to determine uncertainties due to galaxy distribution.The number of pairs is chosen for its similarity in sample size to the selected pair sample. The 200 simulated pairs have average 1/V$_{max}$ Schechter LF parameters, M$^{*}$= -23.2 $\pm$ 0.2 and $\alpha$= -0.9 $\pm$ 0.2. These M$^{*}$  and  $\alpha$ values are the same, within the uncertainties, as those of the 2MASS galaxy Schechter LF parameters found in Kochanek et al. (2001) and Cole et al. (2001). This indicates that the parent sample is not biased from that of Kochanek et al. (2001). The synthetic pair sample parameters are also within the uncertainties found for the true pair sample.This indicates that the selection criteria have not introduced a bias into our luminosity function.

The LF (Fig. 3-4) does not show significant differences in form or derived Schechter parameters (Table 5) based on environmental density. The isolated and grouped pair subsamples have M$^{*}$  within the uncertainties of the entire pair sample and the LF of Kochanek et al. (2001). The $\alpha$ of the isolated subsample is less steep than that of the grouped subsample. Here the Schechter fit at the less luminous end of the isolated subsample is influenced by a larger variation in $\phi$  than that of the grouped and entire pair samples.  
Isolated pairs outnumber grouped pairs by a factor $\sim$2 according to the Schechter $\phi_{0}$ derived in Table 5.

\subsection{Galaxy Merger Rate}
A galaxy merger rate can be estimated as formalized in Patton et al. (2000) where we can adapt a differential merger rate (DMRF) from the DPFF. We adopt 0.5 $\times$ DPFF as the differential ``merger per paired galaxy function". We can then get the differential merger rate by adopting the mass dependent time scales for mergers from Kitzbichler \& White (2008; KW08). The merger timescales found in KW08 are derived from the Millennium Simulation and are found to be dependent on both mass and redshift and are appropriate for projected pairs. They also indicate that use of constant merger timescales for a projected sample yield merger time underestimates. 
The timescale for each mass bin in our differential mass function is taken as the average of the timescale for all galaxies in that bin where each timescale is calculated from the the stellar mass, redshift, and projected separation (M, z, and r$_{p}$), according to the formula of KW08;
\begin{equation}
T_{merge}=2.2 Gyr (\frac{r_{p}}{50 kpc})(\frac{M}{4*10^{10} h^{-1}M_{\sun}})^{-0.3}(1+\frac{z}{8}),
\eqnum{6}
\end{equation}
for $\Delta$v $<$ 300 km s$^{-1}$, and;
\begin{equation}
T_{merge}=3.2 Gyr (\frac{r_{p}}{50 kpc}) (\frac{M}{4*10^{10} h^{-1}M_{\sun}})^{-0.3}(1+\frac{z}{20}),
\eqnum{7}
\end{equation}
for all other pairs of the sample which fall into the range $\Delta$v $<$ 3000 km s$^{-1}$ of KW08.

These criteria most closely match our sample with the caveat that KW08 allow for mergers among galaxies with a factor 4 difference in mass. Timescales here range from 0.3-1.4 Gyr with decreasing mass having a longer timescale for merger.
Our DMRF is 0.5$\times$DPFF/T$_{merge}$ (Figure 5). Our range of DMRF is then $\sim$0.4-1.8$\times$10$^{-2}$ Gyr$^{-1}$.  The volume merger rates (R$_{V}$) determined from 0.5($\phi/h^{3}$)/T$_{merge}$ yield a range for high mass to low mass of$\sim$7$\times$10$^{-7}$ - 2$\times$10$^{-4}$h$^{3}$ Mpc$^{-3}$mag$^{-1}$ Gyr$^{-1}$ (Figure 6). 
A simple comparison of the R$_{V}$ to that derived by Patton\& Atfield (2008) can be achieved by first adjusting for r-band magnitudes of Patton \& Atfield (2008) and the K$_{s}$-band magnitudes of this work. At the same time Patton \& Atfield (2008) use a constant T$_{merge}$ = 0.5 Gyr as opposed to our mass scaled values. Obri\'{c} et al.(2006) finds an (r-K)=2.6 and therefore the two samples overlap in the -18$<$M$_{r}$$<$-22 range.  In this subset of galaxies their volume merger rate of $\sim$6$\times$10$^{-5}$ h$^{3}$ Mpc$^{-3}$ Gyr$^{-1}$ mag$^{-1}$ agrees quite well with our values for galaxies with magnitudes near M$^{*}$ and also decreases for galaxies brighter than M$^{*}$. We also conclude from these 2MASS LF derived merger rates that over 90\% of mergers occur among galaxies with mass $<$10$^{11}$ M$_{\sun}$.

A comparison of these merger rates can be made to the merger rates in cosmological simulations of Maller et al. (2006). Merger rates are derived from these simulations for redshift range
of z$<$0.5 and therefore these are likely to be higher than those derived from our pair sample which has z$<$0.1. The Maller et al. (2006) merger rates are within a factor of 2 of those in our pair sample (Figure 5), within the uncertainties at masses of 4$\times$10$^{10}$ M$_{\sun}$, despite the larger redshift range of the simulation.
Our R$_{V}$ values can be compared to those of Maller et al. (2006) by summing them over the magnitude bins appropriate to the stellar masses listed in their Table 1 as ``high mass" and ``medium mass" samples at the z=0.1 data point. The pair sample of the 2MASS LF  has ``medium mass" and ``high mass"  integrated volume merger rates of 3.4-4.8$\times$10$^{-5}$ h$^{3}$ Mpc$^{-3}$ Gyr$^{-1}$. These values are within the errors shown in Maller et al. (2006) lowest redshift data points (Figure 6) which includes a larger redshift range, z$<$0.3, than that of our pairs. Patton \& Atfield (2008) also compare their results to Maller et al. (2006) and find a similar agreement on volume merger rates. We note that our pairs that fall into the definition of ``high mass"  in Maller et al. (2006) simulations demonstrate a larger integrated volume merger despite the decreasing rate per magnitude bin due to the larger mass range in the ``high mass" sample which includes 5 of our K$_{s}$-band magnitude bins as opposed to the single magnitude bin of the ``medium mass" sample. Figure 6 includes the Maller et al. (2006) volume merger rates adjusted to a per magnitude scale to compare to our pair data.

\section{Summary and Discussion}
We have identified the current sample of 340 paired galaxies 
covering 5800 sq. degrees. from the 2MASS/SDSS(DR5) catalogs and derived a pair mass function ($K_{s}$ LF) which spans the range from 10$^{9}$ to 10$^{12}$ M$_{\sun}$ ($M_{K}$=-19 to $M_{K}$=-26). By comparing this LF to that of all 2MASS galaxies of Kochanek et al. (2001), we developed a DPFF for the entire set of pairs and two subsamples, an isolated and grouped set of pairs. Overall, the paired sample and the subsets determined by local environment display DPFF which are consistent with the hypothesis that ``close major-merger" pairs are produced in proportion to the number of available galaxies of corresponding mass.  Patton \& Atfield (2008) find the number of companions per galaxy (N$_{c}$) to be near 0.02 for their SDSS galaxy sample and near 0.018 for the Millennium simulation. These overall values of N$_{c}$ as analogous to our derived pair fraction are in agreement with that derived in XSH and within 2$\sigma$ of the fraction derived from this current work. Bell et al. (2006) adjust the XSH pair fraction to coincide with their model parameters and find the value of 1\% $\pm$ 0.5\% is in agreement with the Bell et al. (2006) SDSS pair fraction estimate of 1.1\%. 

The longer T$_{merge}$ for galaxies of low end mass range of our sample, and flat DPFF combine to produce an increased number of mergers per unit volume (R$_V$) with decreasing mass. While the errors in our merger rate calculations are large enough to prevent us from identifying a definitive trend in R$_V$ below 10$^{11}$ M$_{\sun}$, our calculations do show a decrease in R$_V$ above this mass corresponding to a magnitude of $\sim$ M$^{*}$. We conclude from these derived merger rates that over 90\% of mergers occur among galaxies with mass $<$10$^{11}$ M$_{\sun}$.

Xu et al. (2004; XSH) calculated a DPFF for a sample of 19 pairs in the combined 2MASS/2dFGRS catalog of Cole et al. (2001). This DPFF suggested that low mass galaxies were not involved in the pairing process at the same rate as higher mass galaxies. In XSH the pair fraction decreased with decreasing luminosity for masses below 10$^{11}$ instead of the current analysis that the fraction is constant over this range. While the 1/V$_{max}$ method for determining the DPFF are the same as used in this work, evidence for clustering in the 2dFGRS exists between $z$=0.04 and $z$=0.06. In the 1/V$_{max}$ method of determining the LF, clustering can bias the shape of the LF when the maximum finding distance of galaxies corresponds to the distance at which the clustering occurs. A magnitude completeness limit of  $K_{s}$ $<$ 13.5 for XSC makes the maximum finding distances of  $z$=0.04 and $z$=0.06 correspond to galaxies of $M_{K}$=-22 to $M_{K}$= -23. The LF of XSH displays a positive slope in this same absolute magnitude bin range.  Since this range also contains the least luminous galaxies in the pair sample, the shape of the 1/V$_{max}$ Schechter fit is likely unfairly biased on the faint end. This bias is minimized in this work because our ($\rm 1/V_{max}$) LFs extend to $\rm M_K$ magnitudes much fainter than  $\rm M_{K}\sim -22$ due to the larger sample size.
 
\acknowledgments 
This publication makes use of data products from the Two Micron All Sky Survey, which is a joint project of the University of Massachusetts and the Infrared Processing and Analysis Center/California Institute of Technology, funded by the National Aeronautics and Space Administration and the National Science Foundation

This research has made use of the NASA/IPAC Extragalactic Database (NED) and the NASA/ IPAC Infrared Science Archive which are operated by the Jet Propulsion Laboratory, California Institute of Technology, under contract with the National Aeronautics and Space Administration

Funding for the Sloan Digital Sky Survey (SDSS) has been provided by the Alfred P. Sloan Foundation, the Participating Institutions, the National Aeronautics and Space Administration, the National Science Foundation, the U.S. Department of Energy, the Japanese Monbukagakusho, and the Max Planck Society. The SDSS Web site is http://www.sdss.org/.

The SDSS is managed by the Astrophysical Research Consortium (ARC) for the Participating Institutions. The Participating Institutions are The University of Chicago, Fermilab, the Institute for Advanced Study, the Japan Participation Group, The Johns Hopkins University, the Korean Scientist Group, Los Alamos National Laboratory, the Max-Planck-Institute for Astronomy (MPIA), the Max-Planck-Institute for Astrophysics (MPA), New Mexico State University, University of Pittsburgh, University of Portsmouth, Princeton University, the United States Naval Observatory, and the University of Washington.

\clearpage
    		
 \begin{deluxetable}{lrll} 
\tablecolumns{4} 
\tablewidth{0pt} 
\tablecaption{Galaxies with Newly Acquired Redshifts.} 
\tablehead{ \colhead{$\alpha$(2000)}&\colhead{$\delta$(2000)}&\colhead{z}&\colhead{Paired?}\\ 
\colhead{}&\colhead{}&\colhead{(km s$^{-1}$)}&\colhead{}
 }
\startdata 
165.4320&	57.34277&0.048&Paired\\
171.3210&	2.4502&0.049&Paired\\
178.6676&	49.3102&0.054&Paired\\
193.2089&	46.7576&0.061&Paired\\
204.3182	   &      45.2504 &0.061&Paired\\
210.7016	  &39.1268&0.068&No\\
211.4617	&65.7165&0.031&Paired\\
213.6322&	1.7297&0.053&Paired\\
215.9372&	6.6011&0.050&Paired\\
217.0412&	-1.6731&0.118&No\\
228.6957&	4.0661&0.038&Paired\\
 230.9070&	37.8176 &0.023&Paired\\
 233.0476	     &    58.9080 & 0.070&Paired\\
 238.1414&	46.3401&0.061&Paired\\
 239.8566	    &      2.9381 &0.041&Paired\\
  239.6563&	32.4605 &0.049&Paired\\
  240.5160&	26.9610&0.104&$\Delta$m$>$1\\
  240.7290	&36.3522&0.068&$\Delta$m$>$1\\
  241.2719&	40.0913&0.117&No\\
  243.7259&	37.1871&0.058&Paired\\
  249.3647	&46.8350 &0.057&Paired\\
  255.3437	&20.3069 &0.059&No\\

     \enddata 
\end{deluxetable}

\begin{deluxetable}{lcccrcc} 
\tablecolumns{8} 
\tablewidth{0pt} 
\tablecaption{K-Band LF for All Paired Galaxies and Differential Pair Fraction.} 
\tablehead{ \colhead{$M_{K}- 5 log h$}&\colhead{log($M_{stars}/h^{-2}$)}&\colhead{log($\phi/h^{3}$)}&\colhead{log(Error $\phi$)}&\colhead{}&\colhead{Pair Fraction}&\colhead{}\\ 
\colhead{(mag)}&\colhead{(M$_{\sun}$)}&\colhead{(Mpc$^{-3}$mag$^{-1}$)}&\colhead{}&\colhead{N}&\colhead{(\%)}&\colhead{Error}
 }
\startdata 
  -17.75\ldots  &    8.64 & -2.84   &  -2.84    &  1     &   \ldots   &    \ldots\\
  -18.25\ldots     & 8.84 &-2.84   &  -2.84    & 1   & \ldots   &  \ldots\\
  -19.25\ldots      &9.24& -3.62   &    -3.77   &  2     & 1.10  &  1.27\\
  -19.75\ldots      &9.44&  -3.65  &   -3.78    & 2    &1.13	&  1.28\\
  -20.25\ldots      &9.64&   -3.55  &   -3.85    &  5    &1.16	&  1.45\\
  -20.75\ldots      &9.841&  -3.91  &   -4.18    &  4    &1.19	& 0.57\\
  -21.25 \ldots     &10.04&  -4.09   &  -4.45   &  6    &1.22	& 0.58\\
  -21.75 \ldots     &10.24 & -3.95   &  -4.52    & 19   &1.25	& 0.32\\
  -22.25\ldots      &10.44 & -4.23   &  -4.82    & 25    &1.27	& 0.60\\
  -22.75\ldots      &10.64 & -4.03   &  -4.84    & 51    &1.30	& 0.26\\
  -23.25 \ldots     &10.84  &-4.19   &  -5.08    & 77   &1.32	& 0.21\\
  -23.75\ldots      &11.04 & -4.46   &  -5.33    & 66    &1.34	& 0.18\\
  -24.25 \ldots     &11.24 & -4.76   &  -5.61    & 56    &1.35	& 0.32\\
  -24.75 \ldots     &11.44  &-5.59   &  -6.19    & 17   &1.34	& 0.50\\
  -25.25\ldots      &11.64 & -6.27   &  -6.68    &  7   &1.31	& 0.62\\
  -25.75 \ldots     &11.84  &-7.40   &  -7.41    &  1    &\ldots	&  \ldots\\
 
     \enddata 
\end{deluxetable}

\begin{deluxetable}{lcccrcc} 
\tablecolumns{8} 
\tablewidth{0pt} 
\tablecaption{K-Band LF for Isolated Paired Galaxies and Differential Pair Fraction.} 
\tablehead{ \colhead{$M_{K}- 5 log h$}&\colhead{log($M_{stars}/h^{-2}$)}&\colhead{log($\phi/h^{3}$)}&\colhead{log(Error $\phi$)}&\colhead{}&\colhead{Pair Fraction}&\colhead{}\\ 
\colhead{(mag)}&\colhead{(M$_{\sun}$)}&\colhead{(Mpc$^{-3}$mag$^{-1}$)}&\colhead{}&\colhead{N}&\colhead{(\%)}&\colhead{Error}
 }
\startdata 
     -17.75\ldots    &8.64 & -2.84  &-2.84 &    1& \ldots&\ldots\\
     -18.25\ldots    & 8.84 & -2.84  &-2.84  & 1    & \ldots   &  \ldots\\
     -19.25\ldots    &9.24&  -4.02  &   -4.02   &  1  &\dots	&  \dots\\     
     -19.75\ldots	 &9.44 &-4.08  &-4.08 &    1&  \ldots&\ldots\\
     -20.25\ldots   &9.64 & -3.65 & -3.87&   3  & 0.43   & 1.64\\
     -20.75\ldots	 &9.84&  -4.34  &-4.34 &   1&\ldots   &\ldots\\
     -21.25\ldots	 &10.04& -4.78  &-4.93 &    2&0.55  & 0.42\\
     -21.75\ldots	 &10.24& -4.37  &-4.77 &    8&0.62&    0.25\\
     -22.25 \ldots	 &10.44& -4.47  &-4.96 &   14&0.69  &  0.31\\
     -22.75 \ldots	 &10.64& -4.28  &-4.97 &   29&0.76  &  0.17\\
     -23.25\ldots	 &10.84& -4.39 &-5.18 &   50&0.82  &  0.17\\
     -23.75\ldots	 &11.04& -4.63 &-5.42  &  44&0.87 &   0.15\\
     -24.25 \ldots	 &11.24&-4.96  &-5.71  &  36&0.88 &   0.21\\
     -24.75\ldots	 &11.44& -5.81  &-6.29 &    10& 0.84&   0.35\\
     -25.25 \ldots	 &11.64&-6.53  &-6.81 &    4& 0.72&  0.45\\
     -25.75 \ldots	 &11.84& -7.40 & -7.41 &    1& \ldots&   \ldots\\
     	    
  \enddata 
\end{deluxetable}

 \begin{deluxetable}{lcccrcc} 
\tablecolumns{8} 
\tablewidth{0pt} 
\tablecaption{K-Band LF for Paired Galaxies in Groups/Clusters and Differential Pair Fraction.} 
\tablehead{\colhead{$M_{K}- 5 log h$}&\colhead{log($M_{stars}/h^{-2}$)}&\colhead{log($\phi/h^{3}$)}&\colhead{log(Error $\phi$)}&\colhead{}&\colhead{Pair Fraction}&\colhead{}\\ 
\colhead{(mag)}&\colhead{(M$_{\sun}$)}&\colhead{(Mpc$^{-3}$mag$^{-1}$)}&\colhead{}&\colhead{N}&\colhead{(\%)}&\colhead{Error}
 }
\startdata   
   -17.75\ldots        &8.64 & \ldots &\ldots &    0& \ldots&\ldots\\
  -18.25\ldots     &8.84&   \ldots      &    \ldots&  0&\ldots& \ldots\\
  -19.25\ldots     &9.24&   -3.85   &      -3.85 &  1&\ldots & \ldots\\
  -19.75\ldots     &9.44&  -3.85 &      -3.85 &  1&\ldots&\ldots\\
  -20.25\ldots     &9.64&  -4.26 & -4.38 &            2 &0.61        &0.37 \\
  -20.75\ldots     &9.84&   -4.10       &    -4.32 &  3&0.59&  0.39\\
  -21.25\ldots     &10.04 & -4.19       &    -4.47 &  4&0.57& 0.30 \\
  -21.75\ldots     &10.24 & -4.15     &      -4.60 & 11&0.56&  0.30\\
  -22.25\ldots     &10.44 & -4.58   &      -5.00 & 11&0.54&  0.26\\
  -22.75\ldots     &10.64  &-4.38       &    -5.01 & 22&0.53&  0.19\\
  -23.25\ldots     &10.84 & -4.63       &    -5.29 & 27&0.51&  0.11\\
  -23.75\ldots     &11.04 & -4.95      &    -5.59 & 22&0.50&  0.12\\
  -24.25\ldots     &11.24  &-5.22       &    -5.84 & 20&0.49&  0.14\\
  -24.75\ldots     &11.44 & -6.00      &    -6.42 &  7&0.48&  0.19\\
  -25.25\ldots     &11.64  &-6.63       &    -6.86 &  3&0.48&0.43\\
  -25.75 \ldots	 &11.84& \ldots & \ldots &    0& \ldots &  \ldots\\
\enddata 
\end{deluxetable}

\begin{deluxetable}{lcccccc} 
\tablecolumns{8} 
\tablewidth{0pt} 
\tablecaption{Schechter Function Parameters of Paired Galaxy LF.} 
\tablehead{\colhead{Sample}&\colhead{$\alpha$}&\colhead{Error}&\colhead{$M^{*}- 5 log h$}&\colhead{Error}&\colhead{log($\phi_{0}/h^{3}$)}&\colhead{Error}
 }
\startdata 
All Pairs&-1.03&0.09&-23.36&0.09&-3.84& 0.06\\ 
Isolated Pairs&-0.8&0.1&-23.3&0.1&-4.00&0.07\\
Grouped Pairs&-1.2&0.2&-23.4&0.1&-4.27&0.09 \\
Synthetic Pairs&-0.9&0.2&-23.2&0.2&-4.57&0.10 \\
Kochanek et al.&-1.09&0.06&-23.39&0.05&-1.94&0.10 \\
\enddata 
\end{deluxetable}

 	\clearpage

\begin{figure}
    \epsscale{0.85}
    \plotone{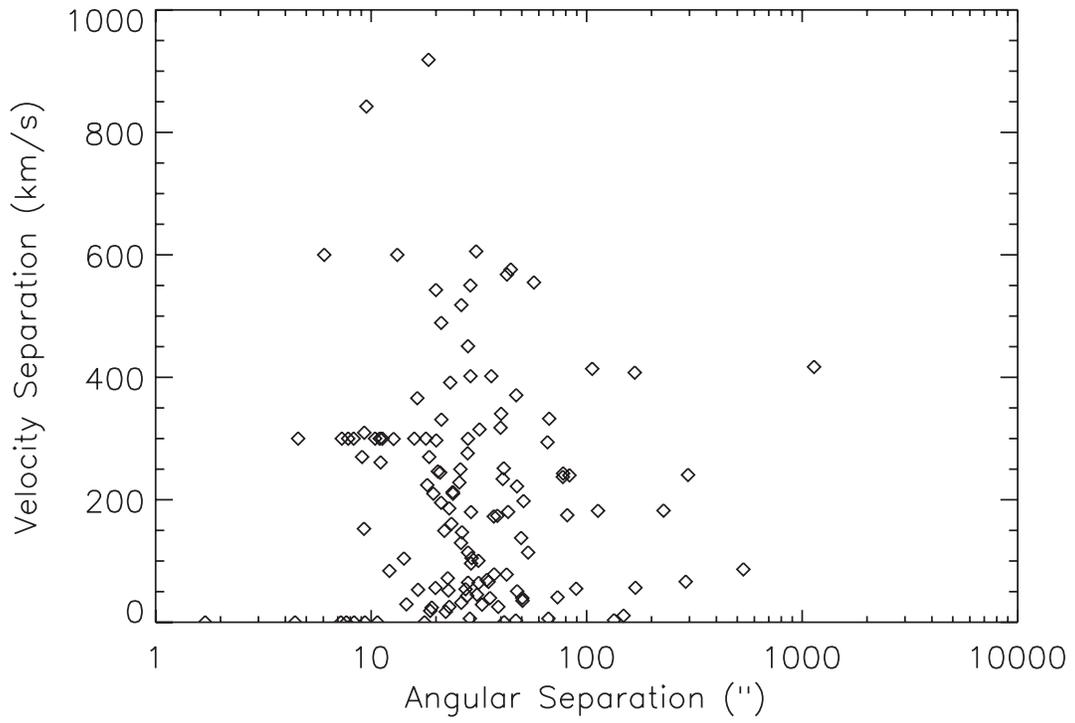}
    \caption{Velocity difference vs. Angular Separation for the sample of pairs with two known redshifts.}
\end{figure}

\begin{figure}
    \epsscale{0.85}
    \plotone{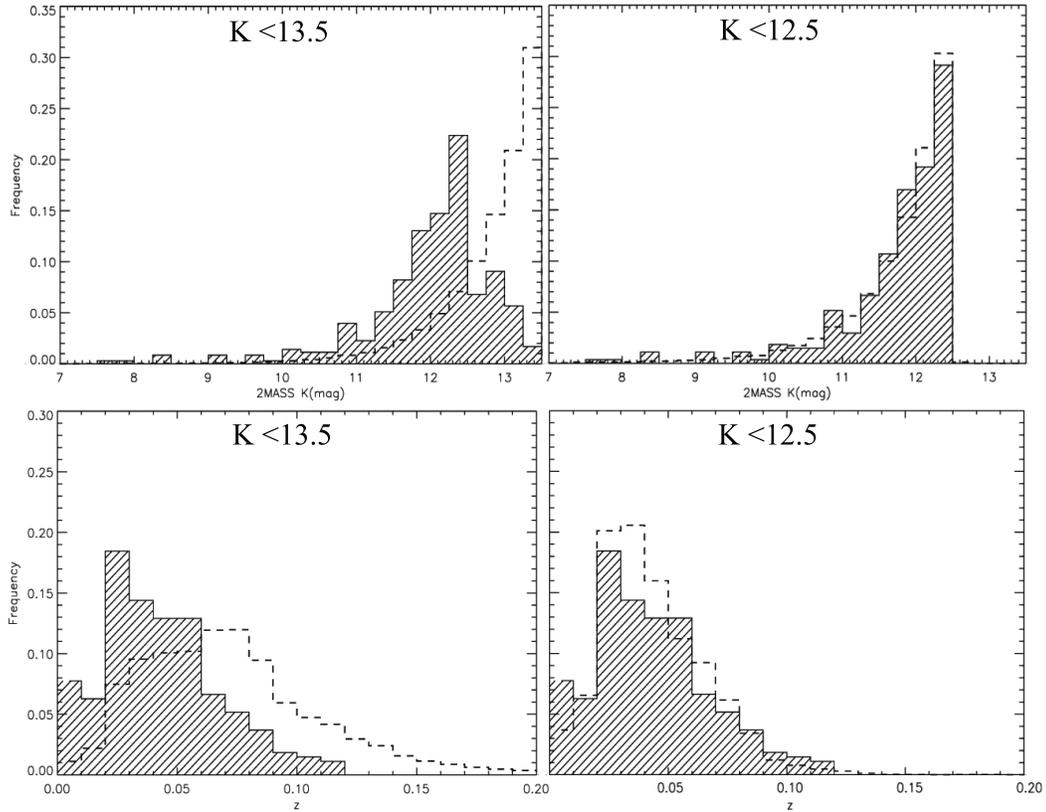}
    \caption{$K_{s}$ and redshift relative distributions for the parent sample (dashed histograms) and the pair sample (shaded histograms). Left Panels represent the entire parent and pair sample to $K_{s}$ $<$ 13.5 while right panels represent the portion of the samples with $K_{s}$ $<$ 12.5.}
\end{figure}

\begin{figure}
    \epsscale{0.85}
    \plotone{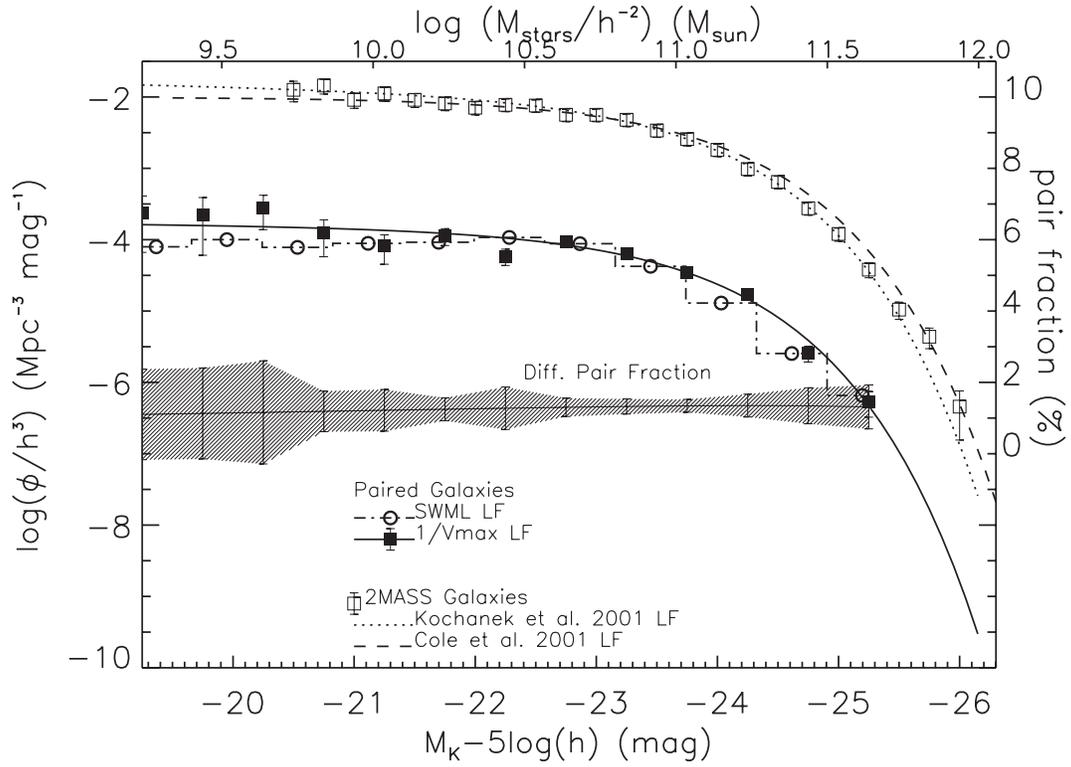}
   \caption{$K_{s}$ Luminosity Functions and stellar mass functions and differential pair fraction (right coordinates).Lines as labeled in the figure are 1/V$_{max}$ Schechter function fit of the paired galaxies, and SWML LF of the paired galaxies, LF of the 2MASS galaxies by Kochanek et al. (2001;dotted), and of the 2MASS galaxies by Cole et al.(2001; dashed). The shaded area represents the differential pair fractions and the errors.}
\end{figure}

\begin{figure}
    \epsscale{0.85}
    \plotone{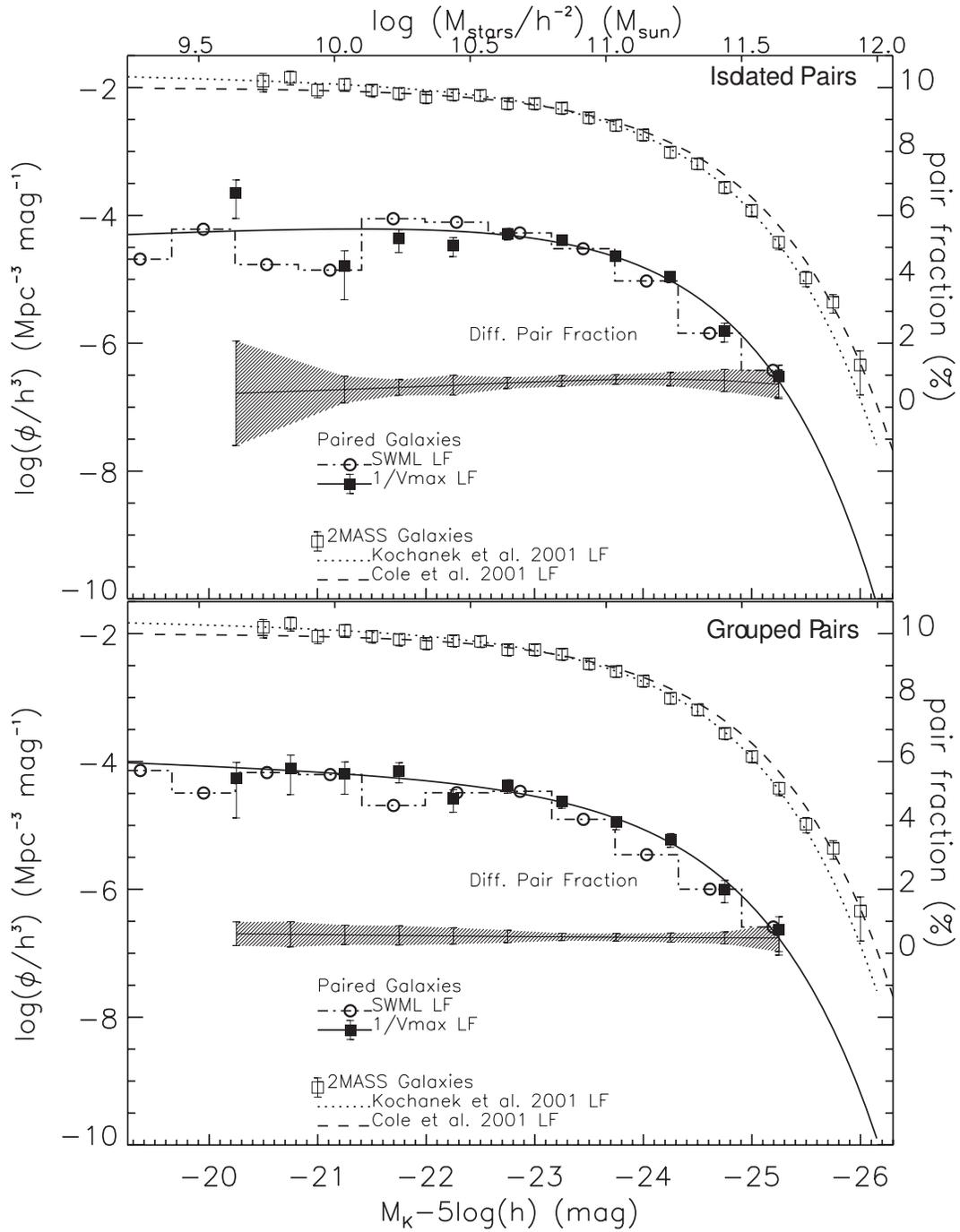}
    \caption{Same as Figure 3 with the isolated and grouped pair samples substituted for the entire pair sample.}
\end{figure}

\begin{figure}
    \epsscale{0.85}
    \plotone{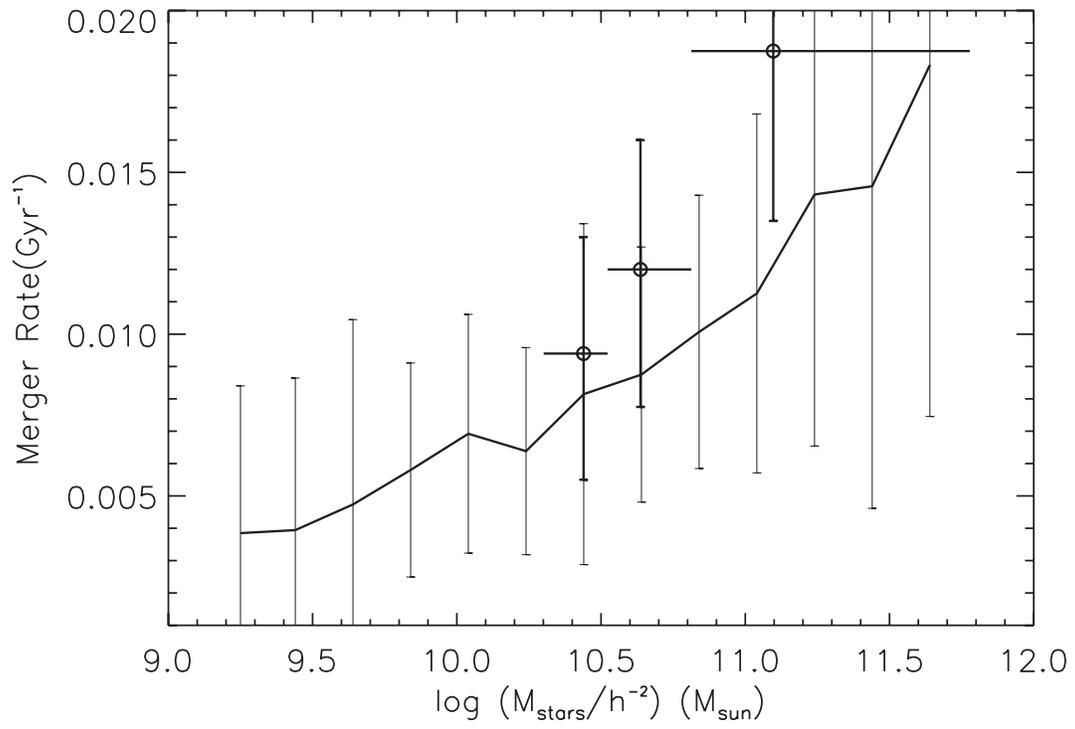}
   \caption{The merger rate (Gyr$^{-1}$) as a function of log(stellar mass) for the pair sample. Open circles are merger rates from the simulations of Maller et al.(2006).}
\end{figure}   
   
   \begin{figure}
    \epsscale{0.85}
    \plotone{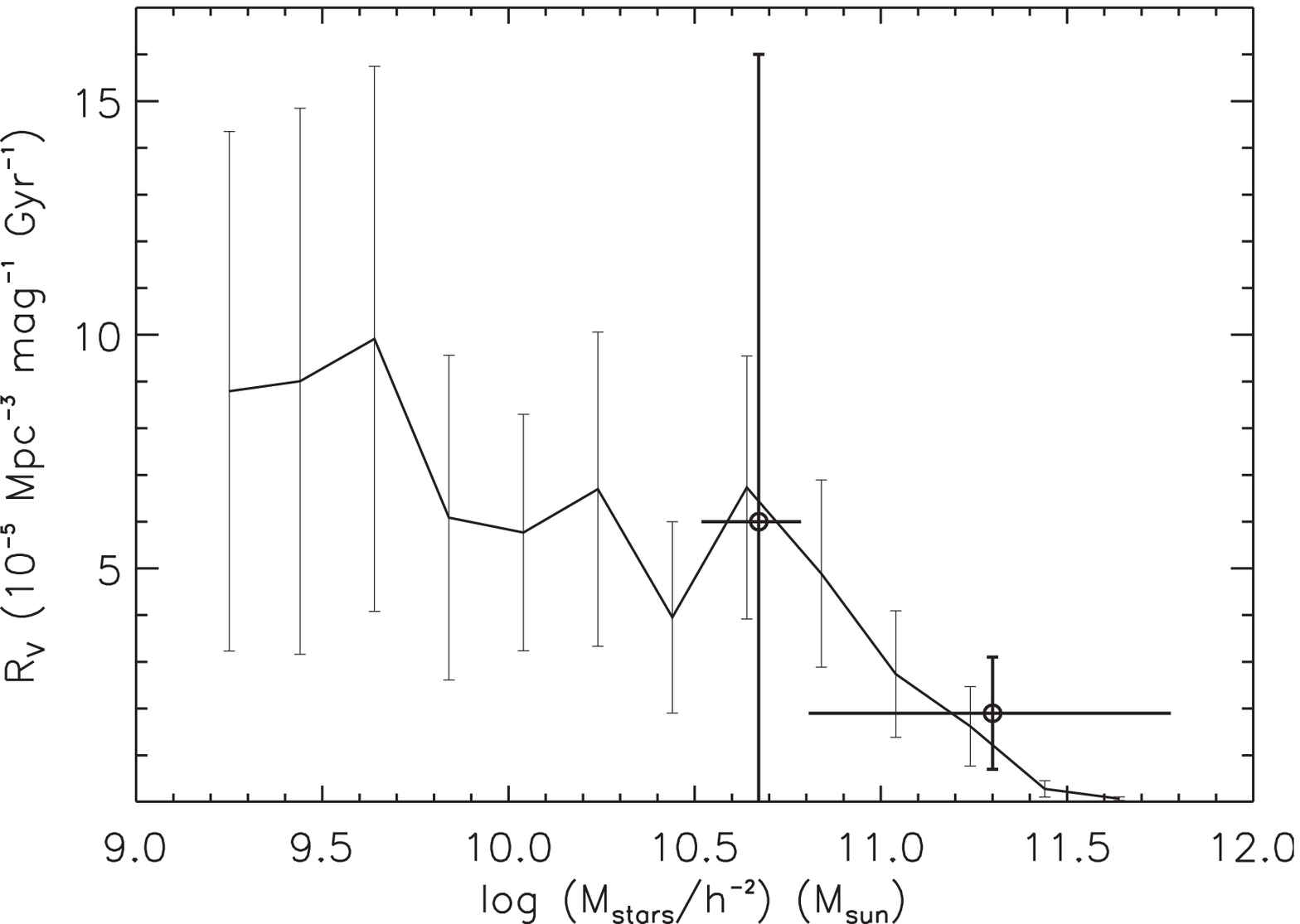}
   \caption{The volume merger rate (10$^{-5}$ h$^{3}$ Mpc$^{-3}$ Gyr$^{-1}$) as a function of log(stellar mass) for the pair sample. Open circles are volume merger rates from the simulations of Maller et al. (2006) after adjusting to a per magnitude scale based on the included mass range of each Maller et al. (2006), z=0.1, data point.}
\end{figure}


\begin{references}
 \reference{}
Adelman-McCarthy, J., et al. 2007, ApJS, 172, 634
 \reference{}
Barnes, J. 1990, Nature, 344, 379
\reference{}
Bell, E. F., et al. 2005, \apj, 625, 23
\reference{}
Bell, E. F., Phleps, S., Somerville, R. S., Wolf, C., Borch, A., Meisenheimer, K. 2006, \apj, 652, 270
\reference{}
Benson, A. J., et al. 2002, MNRAS, 333, 156
\reference{}
Blanton, M. R., et al. 2003a, \aj, 125, 2276
\reference{}
Blanton, M. R., et al. 2003b, \aj, 125, 2348
\reference{}
Brinchmann, J. et al. 1998, \apj, 499, 112
\reference{}
Burkey, J. M., et al. 1994, \apj, 429, L13
\reference{}
Carlberg,R. G., Pritchet, C. J., \& Infante, L. 1994,\apj, 435, 54
\reference{}
Cole, S., et al. 2001, \mnras, 326, 255
\reference{}
Conselice, C. J., Bershady, M. A., Dickinson, M., \& Papovich, C. 2003, AJ, 126, 1183
\reference{}
Daysra, K. M., et al. 2006, \apj, 638, 745
\reference{}
Domingue, D. L., Sulentic, J. W., Durbala, A. 2005, \aj, 129, 2579
\reference{}
Domingue, D. L., Sulentic, J. W., Xu, C., Mazzarella, J., Gao, Y., Rampazzo, R. 2003, \aj, 125, 555
\reference{}
Efstathiou, G., Ellis. R. S., \& Peterson, B. A. 1988, \mnras, 232, 431
\reference{}
Faber, S. M., et al. 2005, astro-ph/0506044
\reference{}
Franceschini, A., Hasinger, G., Miyaji, T., \& Malquori, D. 1999, MNRAS, 310, L5
\reference{}
Geller, M. J., Kenyon, S. J., Barton, E. J., Jarrett, T. H., Kewley, L. J. 2006, \aj,132, 2243
\reference{}
Holmberg, E. 1958, Medd. Lunds Astron. Obs. Ser. II, 136, 1
\reference{}
Jarrett, T. H., Chester, T., Cutri, R., Schneider, S., Skrutskie, M., \& Huchra, J. P. 2000, \aj, 119, 2498
\reference{}
Junqueira, S., de Mello, D. F., Infante, L. 1998, A\&AS, 129, 69
\reference{}
Kartaltepe, J. S., et al. 2007, ApJS, 172, 320
\reference{}
Kauffmann, G., White, S., \& Guiderdoni, B. 1993, MNRAS, 264, 201
\reference{}
Kennicutt, R. C., Jr. 1998, ARA\&A, 36, 189
\reference{}
Kitzbichler, M. G. \& White, S. D. M. 2008, arXiv:0804.1965v1
\reference{}
Kochanek, C. S., et al. 2001, \apj, 560, 566
\reference{}
Kormendy, J., \& Sanders, D. B. 1992, \apj, 390, L53
\reference{}
Le F\'{e}vre, O., et al. 2000, MNRAS, 311, 565
\reference{}
Lin, H., Yee, H. K. C., Carlberg, R. G., \& Ellingson, E. 1997, \apj, 475, 494
\reference{}
Lintott, C. J., et al. 2008, MNRAS, 389, 1179
\reference{}
Maller, A. H., et al. 2006, \apj, 647, 763
\reference{}
Melbourne, J., Koo, D.C., Le Floc'h, E. 2005, \apj, 632, L65
\reference{}
Mould, J. R., et al. 2000, \apj, 529, 786
\reference{}
Obri\'{c}, M. et al. 2006, \mnras, 370, 1677
\reference{}
Patton, D. R., et al. 1997,\apj, 475, 29
\reference{}
Patton, D.R., et al. 2000, \apj, 536, 153
\reference{}
Patton, D.R., et al. 2002, \apj, 565, 208
\reference{}
Patton, D.R. \& Atfield, J. E. 2008, \apj, 685, 235
\reference{}
Schechter, P. 1976, \apj, 203, 297
\reference{}
Schmidt, M. 1968, \apj, 151, 393
\reference{}
Schweizer, F. 1982, \apj, 252, 455
\reference{}
Shimasaku, K. 2001, \aj, 122, 1238
\reference{} 
\reference{}
Strateva, I. et al. 2001, \aj, 122, 1861
\reference{} 
Woods, D., Fahlman, G. G., \& Richer, H. B. 1995, \apj, 454, 32
\reference{}
Woods, D. F., Geller, M. J., \& Barton, E. J. 2006, \aj, 132, 197
\reference{} 
Wu, W., \& Keel, W. C. 1998, \aj, 116, 1513
\reference{}
Yee, H. K. C., \& Ellington, E. 1995, \apj, 445, 37
\reference{}
Zepf, S. E., \& Koo, D. C. 1989, \apj, 337, 34
\reference{}
Xu, C. \& Sulentic, J. W. 1991, \apj, 374, 407
\reference{}
Xu, C. K., Sun, Y. C., \& He, X. T. 2004, \apjl, 603, L73


 \end{references}
 \end{document}